\documentclass[authoryear,10pt]{elsarticle}

\usepackage{natbib}
\usepackage{epsfig}
\usepackage{graphicx}
\usepackage{bm}
\usepackage{amsmath,amssymb,amsthm,amsfonts,mathrsfs,textcomp,mathtools}
\usepackage{url}
\usepackage{multirow}
\usepackage{subfig}
\usepackage{aas_macros}
\usepackage[shortlabels]{enumitem}

\usepackage{hyperref,fullpage}

\makeatletter
\def\namedlabel#1#2{\begingroup
    #2%
    \def\@currentlabel{#2}%
    \phantomsection\label{#1}\endgroup
}
\makeatother

\DeclareFontFamily{U}{mathx}{\hyphenchar\font45}
\DeclareFontShape{U}{mathx}{m}{n}{<-> mathx10}{}
\DeclareSymbolFont{mathx}{U}{mathx}{m}{n}
\DeclareMathAccent{\widebar}{0}{mathx}{"73}

\theoremstyle{plain}

\let\notORI\not 
\usepackage[mathb]{mathabx}
\let\not\notORI

\usepackage{tikz,pgfplots,ifthen}
\usetikzlibrary{arrows,intersections,backgrounds,decorations.pathreplacing}
\usetikzlibrary{positioning,calc,plotmarks}
\def\centerarcfill[#1](#2)(#3:#4:#5){\draw[#1] (#2) --
  ($(#2)+({#5*cos(#3)},{#5*sin(#3)})$) arc (#3:#4:#5);}

\usepackage{xcolor} 
\definecolor{MBlu}{rgb}{.6,.6,3}
\definecolor{DBlu}{rgb}{.1,.1,.7}

\graphicspath{{figures/}}

\def\R{\mathbb{R}}

\newcommand{\clomon}{{\sc clomon}} 

\begin{document}

\title{Improving impact monitoring through Line Of Variations
  densification}

\author[sds,pi]{A.~Del~Vigna}
\ead{delvigna@spacedys.com}
\author[sds]{F.~Guerra}
\author[iaps,ifac]{G.~B.~Valsecchi}

\address[sds]{Space Dynamics Services s.r.l., via Mario Giuntini,
  Navacchio di Cascina, Pisa, Italy}
\address[pi]{Dipartimento di Matematica, Universit\`a di Pisa, Largo
  Bruno Pontecorvo 5, Pisa, Italy}
\address[iaps]{IAPS-INAF, via Fosso del Cavaliere 100, 00133 Roma,
  Italy}
\address[ifac]{IFAC-CNR, via Madonna del Piano 10, 50019 Sesto
  Fiorentino, Italy}

\begin{small}
    \begin{abstract}
        We propose a densification algorithm to improve the Line Of
        Variations (LOV) method for impact monitoring, which can fail
        when the information is too little, as it may happen in
        difficult cases. The LOV method uses a 1-dimensional sampling
        to explore the uncertainty region of an asteroid. The close
        approaches of the sample orbits are grouped by time and LOV
        index, to form the so-called returns, and each return is
        analysed to search for local minima of the distance from the
        Earth along the LOV. The strong non-linearity of the problem
        causes the occurrence of returns with so few points that a
        successful analysis can be prevented. Our densification
        algorithm tries to convert returns with length at most 3 in
        returns with 5 points, properly adding new points to the
        original return. Due to the complex evolution of the LOV, this
        operation is not necessarily achieved all at once: in this
        case the information about the LOV geometry derived from the
        first attempt is exploited for a further attempt. Finally, we
        present some examples showing that the application of our
        method can have remarkable consequences on impact monitoring
        results, in particular about the completeness of the virtual
        impactors search.
    \end{abstract}
\end{small}
\maketitle

\begin{small}
\end{small}

\section{Introduction}
\label{sec:intro}

The main goal of impact monitoring is to establish whether an
Earth-crossing asteroid could possibly impact our planet. This
activity has to be performed as soon as new asteroids are discovered
or as new observations are added to prior discoveries, and the
resulting information has to be immediately spread to solicit
follow-up. Currently, there are two independent impact monitoring
systems, namely \clomon-2 and Sentry, respectively operating at
SpaceDyS\footnote{\url{http://newton.spacedys.com/neodys2/index.php?pc=4.1}}
and JPL\footnote{\url{http://cneos.jpl.nasa.gov/sentry/}}, and
providing the list of asteroids with a non-zero probability of
collision with the Earth within a century.

Both \clomon-2 and Sentry are based on the LOV method explained in
\cite{milani:clomon2}. The basic idea is to represent the uncertainty
region of the asteroid through the LOV, a curve in the initial orbital
elements space, and to study its dynamical evolution in the
future. Since an analytical way to treat the problem cannot exist, the
LOV is sampled by means of a suitable number of points (see
Section~\ref{sec:LOV}), which are then propagated for 100 years in
order to keep track of all the close approaches with the Earth. A
classical tool to study a close approach is the associated Target
Plane (TP) \citep{valsecchi:resret}. One of the advantages of using
the TP is that it translates the possibility of an impact to a very
simple geometric condition, that is, the intersection of the
trajectory with the TP has to be inside a disk centred in the Earth
and with a suitable radius accounting for gravitational
focusing. Therefore, the strategy of the LOV method is to assess the
possibility of a collision in a given close approach by inferring the
LOV geometry on the corresponding TP from the sampling nodes only. The
reliability of this study thus depends on the number of LOV orbits
that intercept the TP: in particular, if they are a few the
information is sometimes too little to draw any conclusion. In
practice, even if there exists an impacting portion of the LOV, in
such cases its detection can be missed due to the lack of
information. The worst case is represented by singletons, which appear
when the strong non-linearity leaves only a lone point of the original
sampling on the TP (see Section~\ref{sec:motivations}). Although there
exist LOV sampling techniques that in principle guarantee a complete
detection of the impact possibilities with probability down to a
certain level, the previously outlined issues imply that this search
cannot be fully complete in practice. This is a well-known problem, as
pointed out in \cite{delvigna:compl_IM}, where the authors suggested a
densification technique as a possible solution.

We propose a densification method consisting in the addition of new
sample orbits in the LOV portion whose image on the TP is composed of
no more than three points. This threshold is based on the completeness
analysis presented in \cite{delvigna:compl_IM}, where it is shown that
the loss of completeness actually occurs when the number of points on
the TP is 1, 2, or 3. Another peculiarity of our method lies in the
selection of the points to add, which is not a simple task since the
densification is meaningful exactly in the most difficult
cases. Indeed, as we explain in Section~\ref{sec:densification}, the
algorithm is divided in two parts to account for the possible
complexity of the LOV geometry on the TP. We tested our procedure on
real cases, showing the actual improvements it brings about the impact
monitoring problem (see Section~\ref{sec:results}).

\section{The impact monitoring problem}
\label{sec:IM}

In this section we briefly recall the main ideas of the LOV method for
impact monitoring, introducing the notation and the definitions needed
in what follows. The starting point is the solution
$\mathbf{x}^*\in \R^N$ of a non-linear least squares fit, along with
its covariance matrix $\Gamma\coloneq \Gamma(\mathbf{x}^*)$ belonging
to the space of the $N\times N$ real matrices $\mathcal{M}(N;\R)$
\citep{milani:orbdet}. The standard case is $N=6$, when the fit
parameters are the six orbital elements, but $N$ can be also larger
if, additionally, some dynamical parameter is determined, \emph{e.g.},
when accounting for the Yarkovsky effect \citep{vokr:yark}. According
to the probabilistic interpretation of the least squares method, the
nominal solution is surrounded by a set of orbits that are still
compatible with the observational data set, the so-called
\emph{confidence region}. The prediction of possible impacts with the
Earth has to consider all these orbits, up to a fixed confidence level
$\sigma>0$. When the non-linearity is mild, the confidence region can
be approximated by the \emph{confidence ellipsoid}
\[
    Z_{lin}^X(\sigma) \coloneqq \left\{\mathbf{x}\in \R^N \,:\,
    (\mathbf{x}-\mathbf{x}^*)^\top C(\mathbf{x}^*)
    (\mathbf{x}-\mathbf{x}^*) \leq \sigma^2\right\},
\]
where $C(\mathbf{x}^*)\coloneq \Gamma(\mathbf{x}^*)^{-1}$ is the
normal matrix.

\subsection{The LOV method}
\label{sec:LOV}

The purpose of impact monitoring is to scan the confidence region
looking for \emph{Virtual Impactors} (VIs), which are connected
subsets of initial conditions leading to a collision with the
Earth. To this end, the confidence region is sampled by a finite set
of orbits, called \emph{Virtual Asteroids} (VAs). Currently, the
algorithm shared by \clomon-2 and Sentry uses a 1-dimensional sampling
method based on the LOV, which is a smooth line in the orbital
elements space \citep{milani:multsol}. The main advantage of this
approach is that the set of VAs has a geometric structure, that is
they belong to a differentiable curve along which interpolation is
possible. The LOV sampling computation provides a set of orbits
$\{\mathbf{x}(\sigma_i)\}_{i=-M,\,\ldots,\,M}$, where $\sigma_i$ are
the LOV parameters. The next step for impact monitoring consists in
the propagation of each VA in the future \citep{milani:AN10}, commonly
for a time span of 100 years\footnote{Non-gravitational forces can be
  relevant for a reliable impact risk assessment when having a longer
  time horizon for the potential impact search. Currently, there are
  four cases that required the inclusion of the Yarkovsky effect in
  term of hazard assessment: (101955)~Bennu
  \citep{milani:Bennu_impact,chesley:Bennu}, (99942)~Apophis
  \citep{chesley:apophis, giorgini:apophis, vokro:apophis,
    farnocchia:apophis}, (29075)~1950~DA
  \citep{giorgini:29075,farnocchia:29075}, and (410777)~2009~FD
  \citep{spoto:410777,delvigna:410777}.}. As anticipated in the
introduction, to detect the close approaches of a VA we first consider
the TP, which is the plane passing through the Earth centre and
orthogonal to the incoming asymptote of the hyperbola defining the
two-body approximation of the VA trajectory at the time of closest
approach. To avoid geometric complications, we call ``close'' only
those approaches with a distance from the Earth centre of mass not
exceeding some value, commonly fixed to $R_{TP}=0.2$~au, and thus the
TP is actually a disk with radius $R_{TP}$. Lastly, to keep track of a
close approach we define a function $\mathbf{f}:\R^N\rightarrow \R^2$
that maps an orbit $\mathbf{x}$ experiencing a close encounter with
the Earth to a point $\mathbf{y}=(\xi,\zeta)\in \R^2$ on the TP. This
function is the composition between the propagation from the initial
epoch to the closest approach time and the conversion to the TP
coordinates. Actually, inside a given close approach there can be
several local minima of the geocentric distance: the definition of
$\mathbf{f}$ can be extended to each of these minima and consequently,
in general, there is more than one TP trace corresponding to a single
orbit $\mathbf{x}$.

According to \cite{milani:clomon2}, the list of the close encounters
of all the VAs is decomposed into \emph{showers} and
\emph{returns}. In particular, first the close approaches are
clustered by date to obtain the showers and each shower is further
divided in LOV segments with consecutive indices, the returns. It can
happen that there is not a clear clustering in time among the
encounters, causing the presence of very long showers and possibly
multiple occurrences of some VA in the same return. In such cases, a
further decomposition scheme is applied, as described in
\cite{delvigna:compl_IM}, in order to produce returns free of
duplications: this makes the interpolation along the LOV possible
inside a return. Moreover, the algorithm guarantees the completeness
of the decomposition procedure, as proved in the paper.

Each return corresponds to a set of TP traces, that is a sampling of
the projection of the LOV segment associated to the return
itself. Then, each couple of consecutive TP traces
$\mathbf{y}(\sigma_j)$ and $\mathbf{y}(\sigma_{j+1})$ of each return
is analysed to understand the geometry of the LOV in between, as
explained in \cite{milani:clomon2}. More precisely, if $r^2(\sigma)
\coloneqq \xi^2(\sigma)+\zeta^2(\sigma)$ is the squared distance of
the LOV point $\mathbf{x}(\sigma)$ from the Earth centre, the aim is
to find the local minima of $r^2$ in $[\sigma_j,\sigma_{j+1}]$. This
information is provided by the sign and the zeroes of
\[
    f(\sigma) \coloneqq \frac{d r^2}{d\sigma}(\sigma).
\]
Inside a return, only some intervals between consecutive VAs contain a
minimum of $r^2$ and they are identified by a geometric classification
of the TP segment between $\mathbf{y}(\sigma_j)$ and
$\mathbf{y}(\sigma_{j+1})$ (see \cite{milani:clomon2}, Table~1). If
the minimum approach distance can be small, by applying suitable
iterative methods (\emph{regula falsi} and Newton method with bounded
steps) it is possible to determine the minimum distance and the
corresponding LOV orbit $\mathbf{x}(\sigma^*)$, with
$\sigma^*\in (\sigma_j,\sigma_{j+1})$. If the corresponding TP trace
$\mathbf{y}(\sigma^*)$ is inside the Earth impact cross section, then
$\mathbf{x}(\sigma^*)$ is an impactor and, by continuity of
$\mathbf{f}$, there exists a suitable neighbourhood of
$\mathbf{x}(\sigma^*)$ made up of impacting orbits: in this case we
can claim to have found a VI.

The main assumption for the local TP analysis is the \emph{principle
  of simplest geometry}, stating that the geometry of the LOV in each
interval $[\sigma_j,\sigma_{j+1}]$ is as simple as possible. This
translates into two assumptions:
\begin{enumerate}[label={\upshape(SG\arabic*)},wide = 0pt,leftmargin=*]
  \item\label{simpl_geom_1} the function $f$ is defined over the whole
    interval $[\sigma_j,\sigma_{j+1}]$, \emph{i.e.},  the LOV
    projection between the two traces does not exit the TP disk;
  \item\label{simpl_geom_2} the LOV geometry on the TP is the simplest
    one compatible with the known information at the nodes, such as
    the tangent vectors to the LOV projection and the sign of $f$ at
    $\sigma_j$ and $\sigma_{j+1}$.
\end{enumerate}
If the return contains a large number of points, the previous
hypotheses are a good approximation, making the TP analysis easier:
problems can arise when the return is made up of a few points. In
fact, the LOV geometry on the TP can be very wild and difficult to
guess, having the information in the nodes only. Since every close
encounter introduces non-linearity, the LOV behaviour on the TP becomes
progressively more complex as the number of close approaches
increases. Indeed, each close approach typically stretches the LOV on
the TP of the subsequent encounter: the more this effect accumulates,
the less is the number of points gradually found on the TP. The
densification of the LOV sampling allows exactly the treatment of
these cases, as explained in Section~\ref{sec:densification}.

Two important local quantities to measure non-linear effects are the
\emph{stretching} and the \emph{width}, defined to be the square root
of the eigenvalues of the propagated covariance matrix on the TP. From
a geometric point of view, since the differential of $\mathbf{f}$ maps
the confidence ellipsoid $Z_{lin}^X(\sigma)$ onto the confidence
ellipse $Z_{lin}^Y(\sigma)$ on the TP, the stretching and the width
can be seen as the lengths of the semimajor and semiminor axis of
$Z_{lin}^Y(1)$. Another quantity which is meaningful for our problem
is the so-called \emph{stretching along the LOV}, given by
$\left|\frac{d\mathbf{y}}{d\sigma}\right|$, since it
represents the displacement of two TP points as a function of the
difference between the corresponding LOV parameters. Note that the
stretching and the stretching along the LOV coincide in the linear
approximation.

\subsection{Completeness}
\label{sec:compl}

A key concept in impact monitoring is the completeness of the VI
search. The \emph{completeness limit} can be formally defined as the
highest impact probability VI that can escape the detection. However,
since this quantity cannot be explicitly computed, it is replaced
with the \emph{generic completeness limit}, further assuming the full
linearity of $\mathbf{f}$ and that a single point on the TP is enough
to identify a VI \cite{milani:clomon2}. Note that the first hypothesis
implies in particular that the trace of the LOV on the TP is simply a
straight line passing through the Earth centre.

The completeness of the VI search is intimately related to the LOV
sampling method. Indeed, each pair of consecutive LOV points is mapped
to a given TP and the distance between the corresponding traces
depends on the differential of $\mathbf{f}$. In
\cite{delvigna:compl_IM} it is proved that there exists an optimal
step-size choice to achieve a fixed generic completeness level
$IP^*$. In particular, the step-size turns out to be inversely
proportional to the probability density along the LOV\footnote{The
  probability density function along the LOV is the one-dimensional
  Gaussian
  $p(\sigma) = \frac{1}{\sqrt{2\pi}} \exp
  \left({-\frac{\sigma^2}{2}}\right)$, where the mean $\sigma=0$
  corresponds to the nominal orbit $\mathbf{x}^*$.}, resulting in a sampling
that is denser around the nominal solution and more sparse towards the
LOV tips. A maximum value for the step-size is thus used to avoid low
resolution in the tail of the distribution.

\section{Motivations for LOV densification}
\label{sec:motivations}

Since the generic completeness limit is a quantity defined under some
simplified assumptions, there is no guarantee that all the VIs with
probability $IP>IP^*$ are actually found, thus the completeness level
actually achieved has to be measured \emph{a posteriori}. One
possibility is explored in \cite{delvigna:compl_IM}, based on an
empirical law to model the total number of VIs $\mathcal{N}$ as a
function of the impact probability. In particular, the study shows
that $\mathcal{N}$ is proportional to the power-law $IP^{-\frac 23}$
for $IP>IP^*$, and this result is obtained by fitting the contour of
the histogram of $\mathcal{N}$ as a function of $IP$. The difference
between the fitted curve corresponding to the power-law and the
histogram implies that there is a loss of efficiency in the VI search
for impact probabilities slightly above the generic completeness level
$IP^*$, that is the number of actually detected VIs is less than the
expected one. Indeed, the definition of generic completeness assumes
that a single point on the TP is enough to find a VI if it exists, but
in some difficult cases this hypothesis is not satisfied and, even
worse, the missing detection of a VI can occur also when there are a
few points on the TP. A densification of the LOV sampling helps in
revealing the actual geometry of the LOV on the TP, which in turn
should fill the gap between the actual completeness level and the
theoretical value $IP^*$. The scope of our densification technique is
to convert returns with very few points ($\leq 3$) into returns with 5
points, to let the VI search more effective even in these cases.

The strong non-linearity of the map $\mathbf{f}$ can make the
principle of simplest geometry non-reliable. For instance, the
function $f$ could not be defined over the whole interval
$[\sigma_j,\sigma_{j+1}]$: this means that the TP is missed for some
value of $\sigma$, indicating that the two TP points under
consideration do not actually belong to the same return. If at some
stage during the application of an iterative method the TP is missed
for the current value of $\sigma$, the algorithm cannot
proceed. Another possibility is the violation of assumption
\ref{simpl_geom_2} due to the occurrence of multiple local minima of
$r^2$ in the same interval between consecutive LOV points, so that the
iterative method applied converges at most to one of these local
minima. The densification method described in this paper is in
principle able to solve both these problems: indeed, it allows one to
properly cut away from the study a LOV portion that possibly misses
the TP (failure of \ref{simpl_geom_1}) and to separate multiple local
minima in different LOV intervals (failure of \ref{simpl_geom_2}), as
described in Section~\ref{sec:densification}.

The most challenging case is given by singletons, which are returns
consisting of one single point and thus the worst situation in terms
of availability of information. The existence of a lone point on the
TP means that the stretching value is so high that both the previous
and the following VAs miss the TP. Singletons are ignored by Sentry,
whereas \clomon-2 treats them using the Newton method with bounded
steps \cite{milani:clomon2}, but this is not sufficient since it often
leads to an unsuccessful or incomplete VI detection. Our densification
technique, converting in particular singletons in returns with more
points, eases the TP analysis in such demanding cases and in turn
increases the completeness of the VI scan.

\section{Densification procedure}
\label{sec:densification}

As anticipated in Section~\ref{sec:motivations}, our algorithm
consists in densifying returns with 1, 2 or 3 points. The basic idea
is to properly select intermediate LOV indices for the new points to
add to the return and then to apply the standard analysis
(Section~\ref{sec:LOV}) to the densified return.

Formally, let $\mathcal{R}$ be a return of length
$n_{\mathcal{R}}\leq 3$, let $r_1<\dots<r_{n_{\mathcal{R}}}$ be the
(integer) indices of its points, and let $t^{\mathcal{S}}_i$ and
$t^{\mathcal{S}}_f$ be the initial and final times of the shower
$\mathcal{S}$ containing $\mathcal{R}$. If $r$ is one of the real
indices selected for the densification (see
Section~\ref{sec:choice_ind}), the further steps of our method are the
following:
\begin{enumerate}[\upshape(D\arabic*),wide = 0pt,leftmargin=*]
  \item\label{comp_sigma} computation of the LOV parameter $\sigma_r$
    corresponding to $r$;
  \item\label{comp_x} computation of the LOV orbit
    $\mathbf{x}(\sigma_r)$;
  \item\label{comp_y} computation of the TP trace
    $\mathbf{y}(\sigma_r)$, if it exists;
  \item\label{add_y} if step \ref{comp_y} is successful, add
    $\mathbf{y}(\sigma_r)$ to the set of TP traces related
    to~$\mathcal{R}$.
\end{enumerate}
We denote with $\mathfrak{D}_r$ the ensemble of steps
\ref{comp_sigma}-\ref{add_y} above. To accomplish steps
\ref{comp_sigma} and \ref{comp_x}, first we compute the nearest
integer $\bar\jmath$ to $r$, which corresponds to the VA
$\mathbf{x}(\sigma_{\bar\jmath})$ of the initial sampling. Then the
LOV parameter $\sigma_r$ and the LOV orbit $\mathbf{x}(\sigma_r)$ are
determined by applying one step of the standard LOV sampling algorithm
\citep{milani:multsol}. Regarding step \ref{comp_y}, in order to
compute the trace $\mathbf{y}(\sigma_r)$, the corresponding orbit
$\mathbf{x}(\sigma_r)$ needs to have a close approach in the time
interval $[t^{\mathcal{S}}_i,t^{\mathcal{S}}_f]$. If this happens, we
attribute the trace $\mathbf{y}(\sigma_r)$ to the return
$\mathcal{R}$; otherwise, step~\ref{comp_y} fails.

To densify $\mathcal{R}$, we compute $d\coloneq 5-n_{\mathcal{R}}$
real indices, corresponding to the points we would like to add in
order to convert $\mathcal{R}$ into a return with $5$ points. In this
way we obtain a set of $5$ real indices $s_1<\cdots<s_5$, containing
the original return indices $r_1,\,\ldots,\,r_{n_\mathcal{R}}$ and the
new indices $t_1,\,\ldots,\,t_d$. The details for the index selection
are provided in Section~\ref{sec:choice_ind}. Once the indices are
chosen, we apply the procedure $\mathfrak{D}_{t_\ell}$ to add the
point with real index $t_\ell$ to the return, for all
$\ell=1,\,\ldots,\,d$. Since $\mathfrak{D}_{t_\ell}$ may fail, we
associate to every index $t_\ell$ a Boolean value
$\mathfrak{b}(t_\ell)$, which is $1$ or $0$ in case the corresponding
point has been successfully added to the return or not,
respectively. Moreover, for every $i=1,\,\ldots,\,n_{\mathcal{R}}$, we
set $\mathfrak{b}(r_i)=1$ as the point with index $r_i$ already
belonged to $\mathcal{R}$.

At the end of this first phase, some of the $\mathfrak{b}(t_\ell)$ can
be $0$: in case, a second attempt for adding new points is
performed. More precisely, we analyse each couple of consecutive
indices $s_\ell$ and $s_{\ell+1}$, having three possible cases
according to their value of $\mathfrak{b}$.

\begin{itemize}
  \item If $\mathfrak{b}(s_\ell)=\mathfrak{b}(s_{\ell+1})=1$ we assume
    that the LOV trace between these two indices is entirely contained
    in the TP disk. Thus we do not add a new point in between.
  \item If $\mathfrak{b}(s_\ell)=\mathfrak{b}(s_{\ell+1})=0$ we assume
    that the LOV trace is entirely outside the TP disk. Also in this
    case, no further point is considered.
  \item If $\mathfrak{b}(s_\ell)$ and $\mathfrak{b}(s_{\ell+1})$ are
    different, the LOV is partially contained in the TP disk. Without
    loss of generality, suppose that $\mathfrak{b}(s_\ell)=1$ and
    $\mathfrak{b}(s_{\ell+1})=0$. Then we try to add the point with
    index
    \begin{equation}\label{eq:second_fail}
       s_\ell + \frac{s_{\ell+1}-s_\ell}{2^k},
    \end{equation}
    where $k$ is the lowest value in $\{1,\,\ldots,\,k_{max}\}$ for
    which the densification procedure succeeds. For the results of
    this paper we assume $k_{max}=3$.
\end{itemize}
The above assumptions rely on the principle of simplest geometry,
which is more and more reasonable as the intervals become smaller, as
it is the case when densifying.

Summarising, the aim of this second attempt is to add a new
point between each couple of consecutive points with discordant
$\mathfrak{b}$ as resulting from the first attempt. Nevertheless, even
this further attempt may fail in particularly difficult cases, so the
final densified return could not have the maximum possible number of
points. Note that this maximum number is not necessarily $5$, since it
can happen that the first attempt produces a configuration such that
$\mathfrak{b}(s_{\ell-1})=1$, $\mathfrak{b}(s_{\ell})=0$, and
$\mathfrak{b}(s_{\ell+1})=1$, then a full success of the second
attempt adds two points, one with index between $s_{\ell-1}$ and
$s_{\ell}$ and the other one between $s_{\ell}$ and $s_{\ell+1}$.

\subsection{Choice of densification indices}
\label{sec:choice_ind}

The indices for the first attempt of the densification are selected
according to the number of points $n_{\mathcal{R}}$ of the starting
return. As anticipated, the method is applied for returns with
$n_{\mathcal{R}}=1$ (singletons), $n_{\mathcal{R}}=2$ (doubletons) or
$n_{\mathcal{R}}=3$ (tripletons). The densification of a singleton is
achieved by adding some other LOV points around it, whereas for
doubletons and tripletons no points are placed outside the return,
since the analysis before the head and after the tail is already
performed with the Newton method with bounded steps, when necessary.

\begin{figure}[t]
    \begin{center}
        \begin{tikzpicture}[scale=0.7]
            \def\al{30}
            \def\xP{-3.25}
            \def\yP{1.75}
            \def\R{5}
            \def\midal{(\al-90)/2}

            \draw[-latex'] (0,-\R-1.5) -- (0,\R+1.5) node[left] {$\zeta$};
            \draw[-latex'] (-\R-1.5,0) -- (\R+1.5,0) node[below] {$\xi$};

            \coordinate (P) at (\xP,\yP);
            \draw[dashed,thin] (\xP,\yP) node[anchor=south east] {$\mathbf{y}_1$} -- (\xP,0) node[below] {$\xi_1$};
            \draw[dashed,thin] (\xP,\yP) -- (0,\yP) node[right] {$\zeta_1$};

            \draw [very thick, name path=circ] circle (\R);
            \path [domain=-0.65*\R:1.2*\R, samples=5, name path=l] plot({\xP+\x*cos(\al)}, {\yP+\x*sin(\al)});
            \path [name intersections={of = circ and l}];
            \coordinate (A)  at (intersection-1);
            \coordinate (B)  at (intersection-2);
            \draw[thick,gray] (A) -- (B);

            \draw[dotted,thin] (P) -- ($(P)+(0,\R/3)$);
            \centerarcfill[fill=MBlu,draw=blue,thin,opacity=0.2](P)(90:\al:1);
            \path (P) -- ($(P)+({90+\midal}:1)$) node[DBlu,above] {$\alpha_1$};

            \path[name path=smallc,] (P) circle (\R/3);
            \path [name intersections={of = smallc and l}];
            \coordinate (V)  at (intersection-1);
            \draw[thick,DBlu,-latex] (P) -- (V) node[above] {$\mathbf{S}_1$};

            \draw [red,decorate,decoration={brace,amplitude=10pt}] (A) -- (P) node[midway,below,rotate=-\midal,yshift=-0.3 cm] {$\lambda_+$};

            \draw [red,decorate,decoration={brace,amplitude=10pt}] (P) -- (B) node[midway,below,rotate=-\midal,yshift=-0.3 cm] {$\lambda_-$};

            \node[black] at (P) {\large$\bullet$};

        \end{tikzpicture}
    \end{center}
    \caption{Geometrical construction for the densification of a
      singleton.}\label{fig:chord_TP}
\end{figure}
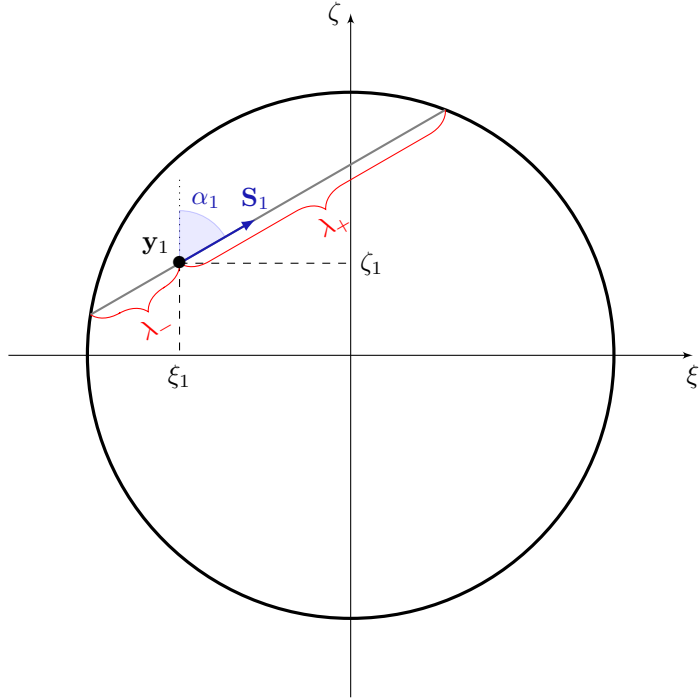

\paragraph{Singletons} The idea behind the densification of a return
with only one point is to exploit the local quantities at that point.
In particular, let $\sigma_{r_1}$ be the LOV parameter of the
singleton, let
$\mathbf{y}_1 \coloneq \mathbf{y}(\sigma_{r_1}) = (\xi_1,\zeta_1)$ be
the corresponding TP trace, let
$\mathbf{S}_1 \coloneq \frac{d\mathbf{y}}{d\sigma}(\sigma_{r_1})$ be
the derivative vector, and let $\alpha_1\in (-\pi,\pi]$ be the angle
from $\mathbf{S}_1$ to the $\zeta$-axis, so that
$\widehat{\mathbf{S}}_1 = (\sin\alpha_1,\cos\alpha_1)$ is the unit
vector of $\mathbf{S}_1$. We consider the chord of the TP disk
parallel to $\widehat{\mathbf{S}}_1$ and passing
through~$\mathbf{y}_1$: this chord is divided into two segments, one
after and one before $\mathbf{y}_1$ according to the direction of
$\widehat{\mathbf{S}}_1$ (see Figure~\ref{fig:chord_TP}). Their length
is respectively
\[
    \lambda_{\pm} = \sqrt{R_{TP}^2 - (\xi_1\cos\alpha_1 -
    \zeta_1\sin\alpha_1)^2} \pm (\xi_1\sin\alpha_1 +
    \zeta_1\cos\alpha_1).
\]

The lengths $\lambda_\pm$ can be converted into lengths of intervals
in the LOV parameter by dividing them for the local stretching along
the LOV $S_1\coloneqq |\mathbf{S}_1|$. This constitutes just an
estimate, since a full computation would require the knowledge of the
stretching as a function of $\sigma$, whereas we only have the value
$S_1$ in $\sigma_{r_1}$. Furthermore, the previous conversion has to
take into account that the resulting length in $\sigma$ cannot exceed
the local step-size value $\Delta\sigma_{r_1}$: indeed, by definition
of singleton, the neighbouring VAs miss the TP to which $\mathbf{y}_1$
belongs. Hence, we compute the lengths in $\sigma$ as
\[
    \Delta\sigma_\pm \coloneq \min \left\{\frac{\lambda_\pm}{S_1},\,
    \delta\cdot \Delta\sigma_{r_1}\right\}.
\]
For the results of this paper we assume $\delta=0.95$ as a security
factor.

Suppose that $f(\sigma_{r_1})=\frac{dr^2}{d\sigma}(\sigma_{r_1})>0$,
that is, moving along the LOV in positive direction, the distance
between the corresponding TP trace and the Earth centre increases. In
this case we place three points before $r_1$ and one point after. More
precisely, according to the notation introduced in
Section~\ref{sec:densification}, we have $d=4$ and we select as
indices
\[
    t_k = r_1 -
    \frac{4-k}{3}\frac{\Delta\sigma_-}{\Delta\sigma_{r_1}},\ \text{for
    }k=1,\,2,\,3, \quad\text{and}\quad t_4 = r_1 + \frac 12
    \frac{\Delta\sigma_+}{\Delta\sigma_{r_1}}.
\]
The other case, that is
$f(\sigma_{r_1})=\frac{dr^2}{d\sigma}(\sigma_{r_1})<0$, is treated
analogously.

\paragraph{Doubletons} The return is composed of two points with
indices $r_1<r_2$, so that $n_{\mathcal{R}}=2$ and $d=3$. The three
points to add for the densification are placed between the two points
of the return: the selected indices are
\[
    t_k = r_1 + k\frac{r_2-r_1}{4},\ \text{for }k=1,\,2,\,3.
\]

\paragraph{Tripletons} The return is composed of three points with
indices $r_1<r_2<r_3$, so that $n_{\mathcal{R}}=3$ and $d=2$. We place
one point between each couple of consecutive points, that is we select
as indices
\[
    t_1 = \frac{r_1+r_2}{2}\quad\text{and}\quad t_2 =
    \frac{r_2+r_3}{2}.
\]

\subsection{Use of bisection method}
\label{sec:bisec}

During the return analysis, whenever possible, the search of the
zeroes of $f$ is performed on intervals $[\sigma_{h_1}, \sigma_{h_2}]$
such that $f(\sigma_{h_1})f(\sigma_{h_2})<0$, where $h_1<h_2$ are two
real indices. In case the return is not densified, we have $h_2-h_1=1$
since the corresponding LOV points are just consecutive VAs. On the
other hand, when the densification procedure is successfully applied
to the return, we have $h_2-h_1<1$. In the first case we adopt an
accelerated version of the modified \emph{regula falsi}, according to
\cite{milani:clomon2}. In principle that method could also be applied
in the second case, but numerical investigations showed that
convergence is not always guaranteed. Indeed the computation of the
function $f$ is not numerically stable, since it involves the
propagation to the TP and the computation of the eigenvalues of the
propagated covariance matrix on the TP, and this effect is
increasingly amplified as the interval becomes more and more
small. For this reason when $h_2-h_1<1$ we switch to the bisection
method, since it results to be more robust in this case
\citep{conte-deboor}. This of course does not mean that the bisection
algorithm always achieves convergence. Actually, a generic iterative
method applied to this situation may not succeed for two main reasons:
the function $f$ is not defined over the whole interval
$[\sigma_{h_1}, \sigma_{h_2}]$ (see Section~\ref{sec:motivations}) or
the maximum number of iterations is exceeded. Despite the use of the
densification procedure lowers the possibility of failures, there is
no guarantee that they do not occur anymore. Anyway, by numerical
evidence, the bisection method over intervals with length less than
$1$ turns out to be the most effective one.

\section{Results}
\label{sec:results}

The results presented in this section were obtained with the software
AstOD, developed in the framework of ESA SSA-NEO program. The software
covers both the orbit determination (OD) and the impact monitoring
(IM) functionalities: the OD component is operational at the NEO
Coordination Centre (NEOCC\footnote{\url{http://neo.ssa.esa.int/}})
since 2017, whereas the IM component was delivered in Spring 2019.

\subsection{Application of the densification procedure}

We introduced the densification procedure in our software AstOD and we
performed the impact monitoring on a list of objects currently present
in the NEODyS Risk List, as of June 2019. For each selected asteroid
we computed two impactor tables, respectively with and without the
densification.  Here we report the results for two sample objects,
namely 2017~WT$_{28}$ and 2008~JL$_{3}$. For both cases we adopted a
non-linear LOV sampling with the setup
\[
    IP^* = 1\times 10^{-7},\quad \sigma_{max} = 5,\quad
    \Delta\sigma_{max}=0.01.
\]
Moreover, we add a separate section for the special case
(29075)~1950~DA: its impact monitoring is remarkably demanding since
it requires taking into account non-gravitational perturbations in the
long-term orbit propagation, differential corrections, LOV sampling
and TP analysis as well as a careful computation of the impact
probability.

The results of the densification procedure are described in details
for some significant returns. We provide each example with a diagram
to give a quick view of the application of the algorithm. The diagrams
share the following basic graphical conventions:
\begin{itemize}
  \item the indices $s_1,\,\ldots,\,s_5$ corresponding to the first
    attempt of densification are marked with a cross;
  \item the indices corresponding to the second attempt of the
    procedure are marked with a star;
  \item each point successfully added is surrounded by a grey circle;
  \item the indices $r_1,\,\ldots,\,r_{n_{\mathcal{R}}}$ of the
    original LOV sampling are surrounded by a double cray circle;
  \item the location of each VI representative is indicated by a blue
    arrow.
\end{itemize}

\paragraph{Asteroid 2017~WT$_{28}$}
This asteroid is a small ($H=28.1$) NEA of the Aten group, with a
non-negligible chance of impacting the Earth in the next century. The
currently available astrometry is quite limited, consisting of 24
optical observations spanning from November to December 2017. Indeed,
the orbit is not very well-constrained, so that the LOV can extend
very far from the nominal orbit. The chaoticity introduced by
subsequent close approaches causes a complex behaviour of the LOV on
the corresponding TPs. This results in a very large amount of returns
with a few points, causing a large number of application of our
densification procedure. Furthermore, the number of detected VIs is
respectively $201$ with the densification and $181$ without it. The
densification improved the VI search not only because it increased the
number of VIs found, but also because some of the newly discovered VIs
have impact probability above the completeness limit
$IP^*=1\times 10^{-7}$. This is particularly remarkable in light of
the discussion of Section~\ref{sec:motivations}.

\begin{figure}[t!]
    \begin{center}
        \begin{tikzpicture}[scale=25]
            \def\indf{{7.9532,8,8.0826,8.1651,8.2477}}
            \def\inds{{8.2064,7,7}}
            \def\int{278}

            \def\ns{1}
            \def\vi{8.14840}

            \def\succf{{1,1,1,1,0}}
            \def\succs{{1,0,0}}

            \def\nod{{0,1,0,0,0}}
            \def\Rf{0.008}
            \def\Rn{\Rf*1.5}
            \def\Rs{\Rf}

            \foreach \i in {1,...,5}{
              \pgfmathsetmacro\p{\indf[\i-1]}
              \node[below] at (\p,-\Rn) {\scriptsize$\int\pgfmathprintnumber{\p}$};
              \node[below] at (\p,-2*\Rn) {\scriptsize$s_{{\i}}$};
              \pgfmathparse{\nod[\i-1]}
              \edef\n{\pgfmathresult}
              \ifthenelse{\n=1}{\draw[semithick,fill=gray!25] (\p,0) circle (\Rn);}{};
              \pgfmathparse{\succf[\i-1]}
              \edef\q{\pgfmathresult}
              \ifthenelse{\q=1}{\draw[semithick,fill=gray!25] (\p,0) circle (\Rf);}{};
              \node at (\indf[\i-1],0) {$\boldsymbol\times$};
            }

            \foreach \i in {1,...,\ns}{
              \pgfmathsetmacro\pp{\inds[\i-1]}
              \pgfmathparse{\succs[\i-1]}
              \edef\qq{\pgfmathresult}
              \ifthenelse{\qq=1}{\draw[semithick,fill=gray!25] (\inds[\i-1],0) circle (\Rs);}{};
              \node at (\pp,0) {$\boldsymbol\star$};
              \node[below] at (\inds[\i-1],-3*\Rn) {\scriptsize$\int\pgfmathprintnumber{\pp}$};
              \draw[densely dotted]  (\inds[\i-1],-\Rn) --  (\inds[\i-1],-3*\Rn);
            }

            \node[above,DBlu] at (\vi,3*\Rn) {\scriptsize{VI}};
            \draw[thick,DBlu,latex'-]  (\vi,0) --  (\vi,3*\Rn);

            \draw[-latex'] (\indf[0]-0.075,0) -- (\indf[4]+0.075,0) node[below] {$s$};

        \end{tikzpicture}
    \end{center}
    \caption{Graphical representation of the densification procedure
      applied to the $2112$ return of asteroid 2017~WT$_{28}$. The
      original return was a singleton: the first densification attempt
      resulted in three new points (encircled crosses) and the second
      attempt added a further point (encircled star). Clearly, the
      index values depend on the method used for the LOV sampling:
      they are reported here for the sake of completeness, but what
      matters are the relative distances.}\label{fig:2017WT28_4364}
    \vspace{0.5cm}
    \centering
    \includegraphics[scale=0.6]{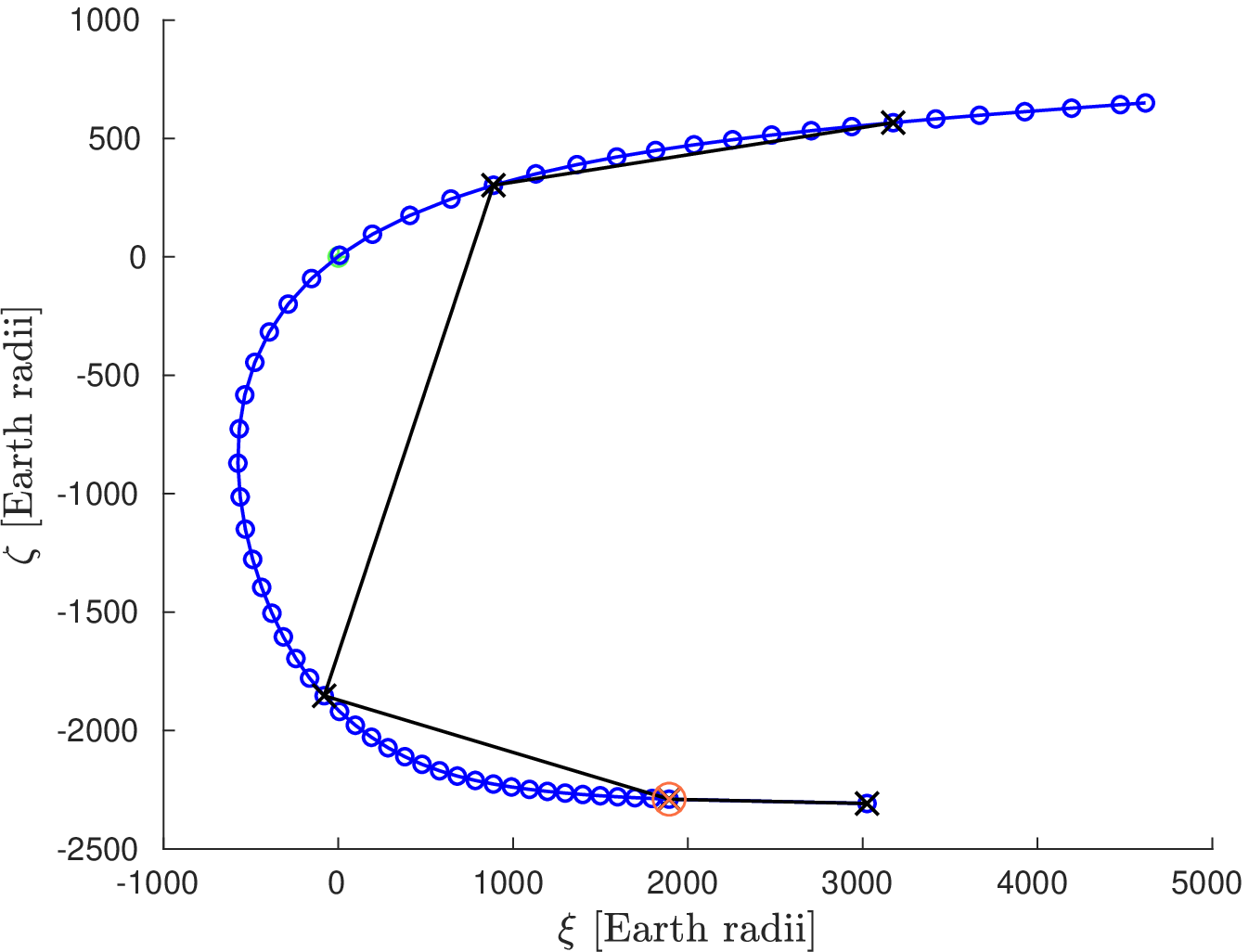}
    \caption{Plot of the LOV projection on the 2112 TP for asteroid
      2017~WT$_{28}$. The five black crosses joined by lines
      represents the densified return resulting from the densification
      procedure. In particular, the orange encircled cross marks the
      original singleton. The actual geometry of the return is
      revealed by the blue circles, which are obtained through a
      refinement of the previous densification. The Earth impact cross
      section is centred in the origin and has radius
      $b_\Earth = 2.7R_\Earth$.}\label{fig:2017WT28_4364_tp}
\end{figure}

\begin{figure}[t!]
    \begin{center}
        \begin{tikzpicture}[scale=8]
            \def\int{401}
            \def\indf{{1,1.25,1.5,1.75,2}}
            \def\inds{{8.2064,7,7}}
            \def\vi{1.06440}

            \def\ns{0}

            \def\succf{{1,1,1,1,1}}
            \def\succs{{0,0,0}}

            \def\nod{{1,0,0,0,1}}
            \def\Rf{0.008*25/8}
            \def\Rn{\Rf*1.5}
            \def\Rs{\Rf}

            \foreach \i in {1,...,5}{
              \pgfmathsetmacro\p{\indf[\i-1]}
              \node[below] at (\p,-\Rn) {\scriptsize$\int\pgfmathprintnumber{\p}$};
              \node[below] at (\p,-2*\Rn) {\scriptsize$s_{{\i}}$};
              \pgfmathparse{\nod[\i-1]}
              \edef\n{\pgfmathresult}
              \ifthenelse{\n=1}{\draw[semithick,fill=gray!25] (\p,0) circle (\Rn);}{};
              \pgfmathparse{\succf[\i-1]}
              \edef\q{\pgfmathresult}
              \ifthenelse{\q=1}{\draw[semithick,fill=gray!25] (\p,0) circle (\Rf);}{};
              \node at (\indf[\i-1],0) {$\boldsymbol\times$};
            }

            \ifthenelse{\ns>0}{\foreach \i in {1,...,\ns}{
		\pgfmathsetmacro\pp{\inds[\i-1]}
		\pgfmathparse{\succs[\i-1]}
		\edef\qq{\pgfmathresult}
		\ifthenelse{\qq=1}{\draw[semithick,fill=gray!25] (\inds[\i-1],0) circle (\Rs);}{};
		\node at (\pp,0) {$\boldsymbol\star$};
		\node[below] at (\inds[\i-1],-3*\Rn) {\scriptsize$\int\pgfmathprintnumber{\pp}$};
		\draw[densely dotted]  (\inds[\i-1],-\Rn) --  (\inds[\i-1],-3*\Rn);
              }}{};

            \node[above,DBlu] at (\vi,3*\Rn) {\scriptsize{VI}};
            \draw[thick,DBlu,latex'-]  (\vi,0) --  (\vi,3*\Rn);
            \def\vi{1.27835}
            \node[above,DBlu] at (\vi,3*\Rn) {\scriptsize{VI}};
            \draw[thick,DBlu,latex'-]  (\vi,0) --  (\vi,3*\Rn);

            \draw[-latex'] (\indf[0]-0.075*25/8,0) -- (\indf[4]+0.075*25/8,0) node[below] {$s$};

        \end{tikzpicture}
    \end{center}
    \caption{Graphical representation of the densification procedure
      applied to the $2114$ return of asteroid 2017~WT$_{28}$. The
      original return was a doubleton: the densification succeeded in
      adding three new points (encircled crosses) all at once in the
      first attempt. Clearly, the index values depend on the method
      used for the LOV sampling: they are reported here for the sake
      of completeness, but what matters are the relative
      distances.}\label{fig:2017WT28_4839}
    \vspace{0.5cm}
    \centering
    \includegraphics[scale=0.6]{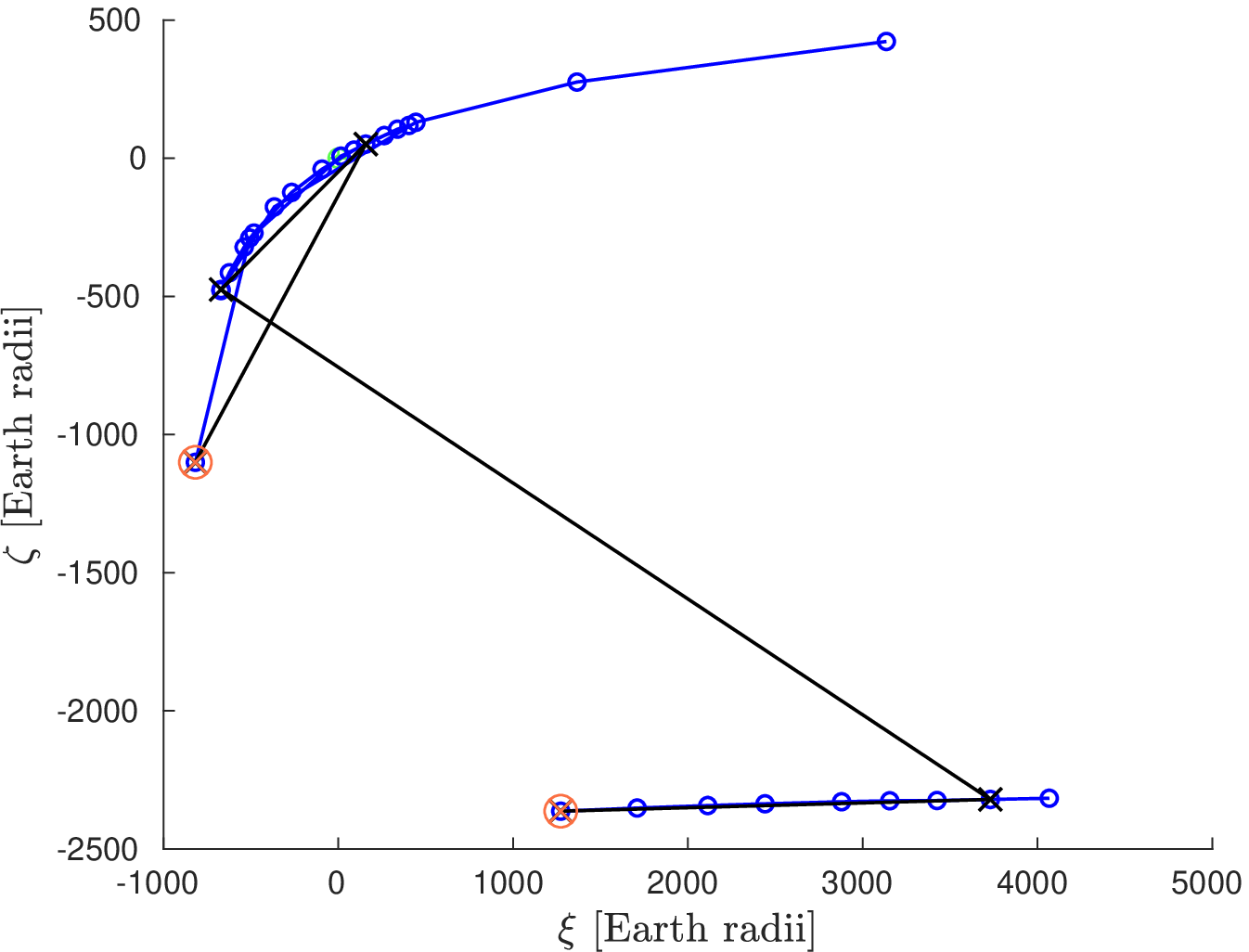}
    \caption{Plot of the 2114 TP for asteroid 2017~WT$_{28}$. The
      orange crosses mark the original doubleton. The five black
      crosses joined by lines are the traces of the densified
      return. The Earth impact cross section is centred in the origin
      and has radius
      $b_\Earth = 2.7R_\Earth$.}\label{fig:2017WT28_4839_tp}
\end{figure}

The first example, represented in Figure~\ref{fig:2017WT28_4364},
corresponds to a return of the 2112 shower of 2017~WT$_{28}$. The
original return is a singleton, that is $n_{\mathcal{R}}=1$, thus the
first phase of the densification tried to add four points. More
precisely, since the distance $r^2$ is decreasing at the singleton
(\emph{i.e.}, at $\sigma_{r_1}$), one of the four points has index
smaller and the other three larger than $r_1$, according to
Section~\ref{sec:choice_ind}. The attempt succeeded for all the points
but the rightmost one, so that in particular $\mathfrak{b}(s_4) = 1$
and $\mathfrak{b}(s_5)=0$. This means that the LOV projection exits
the TP after an index between $s_4$ and $s_5$, due to the rapidly
varying stretching, which is the common situation around a
singleton. Therefore, the second phase of the densification tries to
add a further point with index in between, this time with success,
thus yielding a final densified return with $5$
points. Figure~\ref{fig:2017WT28_4364_tp} shows the original and
densified returns on the 2112 TP. From the shape of the densified
return it is clear that a minimum of the distance along the LOV exists
and that it is located between its third and fourth point. Indeed, the
distance $r^2$ is decreasing at $s_3$ and increasing at $s_4$, that is
its derivative $f(\sigma)$ is negative at the first endpoint and
positive at the second one. Moreover, since the angles $\alpha_3$ and
$\alpha_4$ between the tangent vectors to the LOV projection and the
$\zeta$-axis indicate a large curvature of the LOV, we have an
interrupted return \citep{milani:clomon2}. This is also confirmed by
the plot of Figure~\ref{fig:2017WT28_4364_tp}, showing a further
refined sampling of the return. Additionally, the figure makes clear
the existence of a VI since the refined curve intersects the Earth
disk. To find the minimum of $r^2$ over $[\sigma_{s_3},\sigma_{s_4}]$,
the bisection method was applied, as it always happens for intervals
of a densified return (see Section~\ref{sec:bisec}), yielding the
above-mentioned VI, which is in particular on 2112-11-23.98 and has
impact probability $IP = 1.01\times 10^{-7}$. It is worth mentioning
that the VI would have been missed without using the densification
procedure even if its impact probability is above the generic
completeness level.

The second example is a doubleton in the 2114 shower of the same
asteroid. The application of our procedure converted the doubleton in
a return with $5$ points by adding the three uniformly-spaced points
foreseen in the first attempt (see Section~\ref{sec:choice_ind}), as
the diagram of Figure~\ref{fig:2017WT28_4839}
represents. Figure~\ref{fig:2017WT28_4839_tp} shows the original and
densified returns on the 2114 TP: the LOV geometry unveiled by the
densification procedure was not predictable only from the doubleton
and it is far from being simple since the LOV projection contains at
least two reversals in the portion closer to the Earth. In particular,
this allows the existence of at least two minima of the distance $r^2$
along the LOV, one between the first and the second point and one
between the subsequent pair. The actual LOV geometry can be seen in
Figure~\ref{fig:2017WT28_4839_tp} (blue circles). In the upper part
the LOV passes three times close to the Earth, as suggested by the
position of the points corresponding to $s_2$ and $s_3$, then it
leaves and re-enters the TP on the right side of the plot. Thus, in
this particular case, the resolution foreseen for the doubleton
densification turns out to be not sufficient to reveal the split of
the LOV in the two components. Nevertheless, the densification anyway
improves the knowledge of the LOV geometry for the portions contained
in the TP, which are the relevant ones since the interval containing
the split does not contain any significant VI.

Although also the configuration of the upper LOV portion corresponds
to two interrupted returns, unlike the previous example the derivative
$f$ of $r^2$ assumes the same sign on the endpoints of both the
intervals $[\sigma_{s_1},\sigma_{s_2}]$ and
$[\sigma_{s_2},\sigma_{s_3}]$. Such cases deserve a special analysis,
according to \cite{milani:clomon2} (interrupted failed configuration):
in our example this analysis ends up with the detection of one VI in
each interval, with impact probabilities $IP=8.37\times~10^{-8}$ and
$IP=1.19\times 10^{-7}$, respectively, the first on 2114-11-24.63 and
the other on 2114-11-24.75. Note that the second VI is above the
completeness level and both VIs would not have been found without
applying the densification procedure, as already stressed for the
previous example.

\begin{figure}[h!]
    \begin{center}
        \begin{tikzpicture}[scale=8]
            \def\indf{{59,59.25,59.50,59.75,60}}
            \def\inds{{59.0312,59.6250}}
            \def\int{3}

            \def\ns{2}
            \def\vi{59.85611}

            \def\succf{{1,0,0,1,1}}
            \def\succs{{1,1}}

            \def\nod{{1,0,0,0,1}}
            \def\Rf{0.008*25/8}
            \def\Rn{\Rf*1.5}
            \def\Rs{\Rf}

            \foreach \i in {1,...,5}{
              \pgfmathsetmacro\p{\indf[\i-1]}
              \node[below] at (\p,-\Rn) {\scriptsize$\int\pgfmathprintnumber{\p}$};
              \node[below] at (\p,-2*\Rn) {\scriptsize$s_{{\i}}$};
              \pgfmathparse{\nod[\i-1]}
              \edef\n{\pgfmathresult}
              \ifthenelse{\n=1}{\draw[semithick,fill=gray!25] (\p,0) circle (\Rn);}{};
              \pgfmathparse{\succf[\i-1]}
              \edef\q{\pgfmathresult}
              \ifthenelse{\q=1}{\draw[semithick,fill=gray!25] (\p,0) circle (\Rf);}{};
              \node at (\indf[\i-1],0) {$\boldsymbol\times$};
            }

            \foreach \i in {1,...,\ns}{
              \pgfmathsetmacro\pp{\inds[\i-1]}
              \pgfmathparse{\succs[\i-1]}
              \edef\qq{\pgfmathresult}
              \ifthenelse{\qq=1}{\draw[semithick,fill=gray!25] (\inds[\i-1],0) circle (\Rs);}{};
              \node at (\pp,0) {$\boldsymbol\star$};
              \node[below] at (\inds[\i-1],-3*\Rn) {\scriptsize$\int\pgfmathprintnumber{\pp}$};
              \draw[densely dotted]  (\inds[\i-1],-\Rn) --  (\inds[\i-1],-3*\Rn);
            }
            \node at (59.125,0) {$\boldsymbol\star$};
            \node at (59.0625,0) {$\boldsymbol\star$};

            \node[above,DBlu] at (\vi,3*\Rn) {\scriptsize{VI}};
            \draw[thick,DBlu,latex'-]  (\vi,0) --  (\vi,3*\Rn);

            \draw[-latex'] (\indf[0]-0.075,0) -- (\indf[4]+0.075,0) node[below] {$s$};

        \end{tikzpicture}
    \end{center}
    \caption{Graphical representation of the densification procedure
      applied to the $2109$ return of asteroid 2008~JL$_{3}$. The
      original return was a doubleton: the first densification attempt
      resulted just in one new point out of three (encircled cross),
      whereas the second attempt provided two further points
      (encircled star). Note that the two stars without circle
      indicate that the leftmost point added in the second attempt was
      obtained for $k=3$ in \eqref{eq:second_fail}. Clearly, the index
      values depend on the method used for the LOV sampling: they are
      reported here for the sake of completeness, but what matters are
      the relative distances.}\label{fig:2008JL3}
    \vspace{0.5cm} \centering
    \includegraphics[scale=0.6]{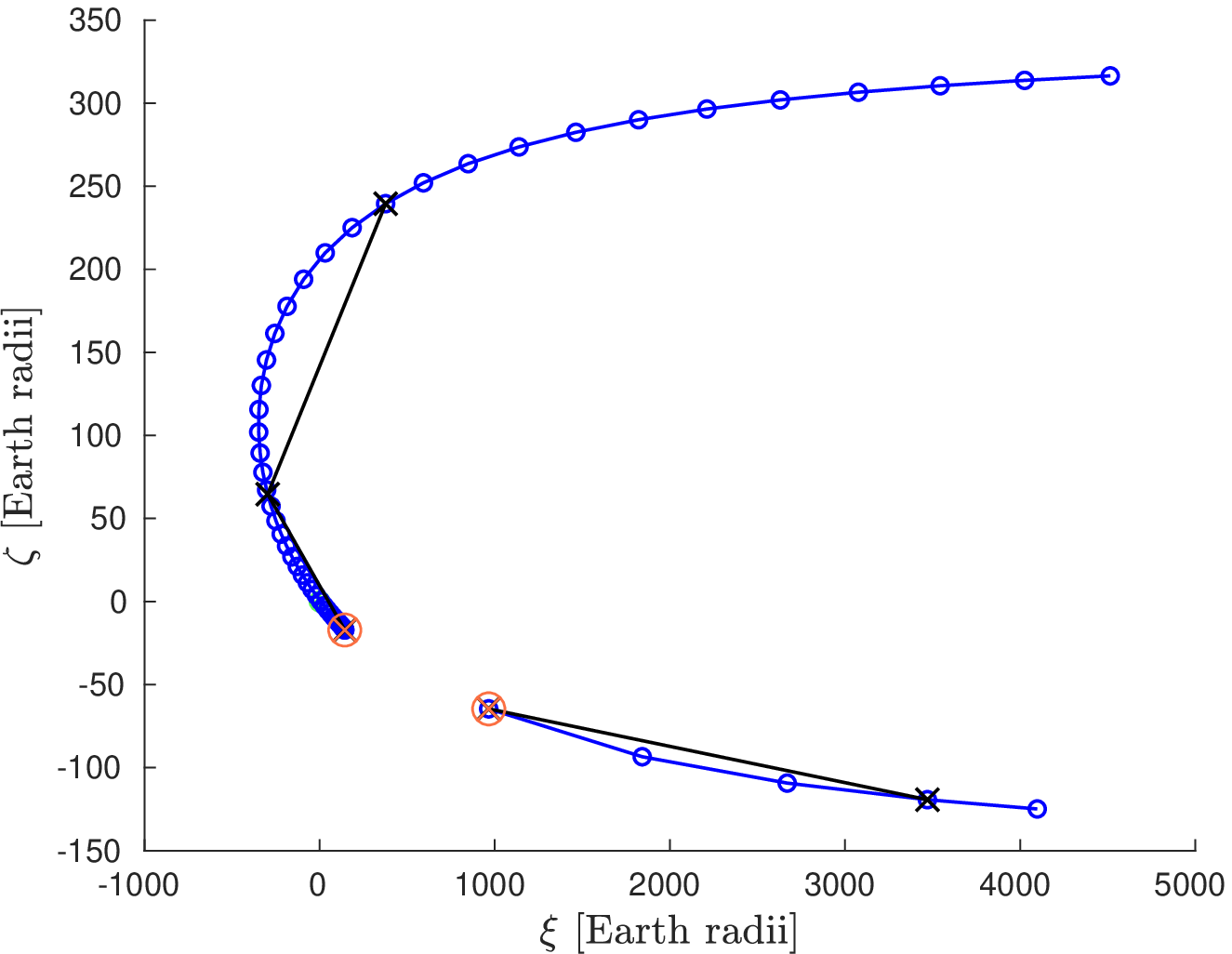}
    \caption{Plot of the LOV projection on the 2109 TP for asteroid
      2008~JL$_{3}$. The five black crosses joined by lines represents
      the densified return resulting from the densification
      procedure. Note that there is an interruption between the second
      and the third point of the densified return, due to the LOV exit
      from the TP. The orange encircled crosses mark the original
      doubleton. The actual geometry of the return is revealed by the
      blue circles, which are obtained through a refinement of the
      previous densification. The Earth impact cross section is
      centred in the origin and has radius
      $b_\Earth = 1.7R_\Earth$.}\label{fig:2008JL3_tp}
\end{figure}

\paragraph{Asteroid 2008~JL$_{3}$}
This asteroid is a NEA of the Apollo group, currently contained in the
upper part of the risk list sorted by Palermo Scale, having
$PS=-3.95$. Its nominal orbit is quite uncertain, due to the short
observational arc which spans only four days. The application of our
method to the 2109 return is a particularly interesting example for
several reasons. The original return is a doubleton, which is
converted in a densified return with three additional points at the
end of the overall procedure, as shown in
Figure~\ref{fig:2008JL3}. Note that the first phase succeeded just for
the last point, so that $\mathfrak{b}(s_1)=1$,
$\mathfrak{b}(s_2)=\mathfrak{b}(s_3)=0$, and
$\mathfrak{b}(s_4)=\mathfrak{b}(s_5)=1$. This means that the LOV is
not entirely contained in the TP between $s_2$ and $s_3$ and in the
second phase of the procedure the LOV shape on the TP is better
understood by adding a point before $s_2$ and a point after $s_3$. The
strong non-linearity around the first point of the original return,
which is confirmed by the failure in $s_2$, makes the addition of the
point between $s_1$ and $s_2$ more difficult: indeed, the goal is
reached at the maximum number of iterations $k_{max}=3$ in
equation~\eqref{eq:second_fail}. The subsequent return analysis
established the existence of a minimum of the distance $r^2$, located
between $s_4$ and $s_5$, and corresponding to an impacting orbit on
2109-04-27.96. The related VI has $IP=2.6\times 10^{-7}$ and would
have been missed without densifying. As in the previous cases, we plot
in Figure~\ref{fig:2008JL3_tp} a refinement of the densified return
sampling to further validate our method: indeed the behaviour of the
corresponding blue curve is well-represented by the five points
resulting from our procedure, which also accounts for the exit of the
LOV from the TP. Moreover, by looking at the two impact monitoring
systems, a remarkable fact comes out: Sentry detects the VI with an
impact probability comparable to ours, whereas NEODyS does not find it
although the impact probability is above the completeness level. This
again shows that the densification can actually improve the overall
efficiency of the system in finding VIs.

\subsection{The special case of (29075)~1950~DA}

\begin{figure}[b!]
    \centering
    \includegraphics[scale=0.6]{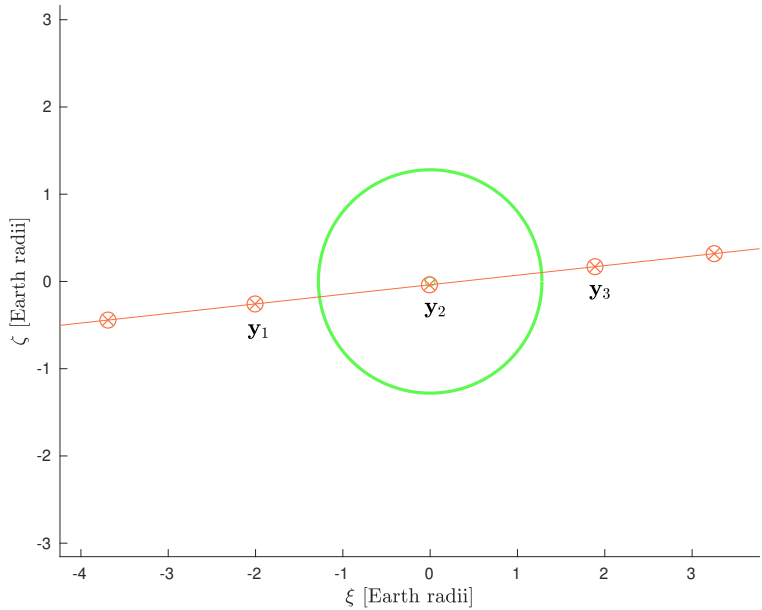}
    \caption{LOV projection on the 2880 TP for asteroid $(29075)$,
      corresponding to the linear LOV with 1200 points per side
      over the interval $|\sigma|\le 3$.}\label{fig:tp_nodens}
\end{figure}

Asteroid (29075)~1950~DA was discovered in 1950 and then lost until
December 2000, when it was recognised to be the object
2000~YK$_{66}$. The current optical observation data set covers a long
arc, from 1950 to 2018. Furthermore, 12 radar observations were added
during the two apparitions of 2001 and 2012. The large extent of the
arc and the availability of radar measurements allow a very precise
orbit determination, with the possibility to fit also the Yarkovsky
parameter $A_2$. The inclusion of the Yarkovsky effect in the
dynamical model is needed to make a reliable hazard assessment for the
2880 possible impact \citep{giorgini:29075,farnocchia:1950DA}. In
particular, the value of $A_2$ computed by AstOD is
$(-5.92\pm 1.25) \times 10^{-15}$~au/d$^2$, which corresponds to a
semimajor axis drift of $da/dt = (-2.60\pm 0.55)\times
10^{-4}$~au/My. This is well-consistent with the $A_2$ estimate
$(-6.03\pm1.25)\times 10^{-15}$~au/d$^2$ of \cite{delvigna:yarko},
obtained with the OrbFit software.

Concerning the impact monitoring of (29075), we considered the
7-dimensional space of the orbital elements and the Yarkovsky
parameter and we adopted the linear approximation of the LOV since the
initial confidence region is small. In particular, a LOV sampling with
1200 points per side over the interval $|\sigma|\le 3$ is sufficient
to detect the 2880 VI. The resulting LOV projection on the 2880 TP is
represented in Figure~\ref{fig:tp_nodens}. There is only one TP trace
inside the Earth impact cross section: let us denote it with
$\mathbf{y}_2$, and let $\mathbf{y}_1,\,\mathbf{y}_3$ be the previous
and subsequent traces, respectively. As it is clear from
Figure~\ref{fig:tp_nodens}, the minimum of the distance from the Earth
centre exists and it is located between $\mathbf{y}_2$ and
$\mathbf{y}_3$. Indeed, the function $f$ has opposite sign at the
interval endpoints and the LOV curvature is negligible, so the
\emph{regula falsi} method easily achieved convergence, yielding a VI
representative.

In general, given a point on the LOV leading to an impact, the impact
probability of the associated VI is usually computed by integrating a
2-dimensional linearised probability density function over the Earth
impact cross section on the TP. When the width is small (few
kilometres) and the impact probability is comparatively high, a more
accurate estimate can be achieved by integrating the 1-dimensional
probability density function $p(\sigma)$ over the preimage under
$\mathbf{f}\circ \mathbf{x}$ of the chord resulting from the
intersection between the LOV trace and the Earth disk. This preimage
is an interval $[\sigma_i,\sigma_f]$ in the LOV parameter space and
its endpoints can be computed once the endpoints of the chord are
known. When $\sigma_i$ and $\sigma_f$ have been determined, the impact
probability can be estimated as
\[
    IP = \frac{1}{\sqrt{2\pi}}\int_{\sigma_i}^{\sigma_f}
    e^{-\frac{\sigma^2}2}d\sigma.
\]
A way to obtain a good approximation of the chord endpoints is to
perform a local densification of the return between $\mathbf{y}_1$ and
$\mathbf{y}_3$ to place a suitable number of points along the
chord. This strategy also allows the numerical computation of the
above integral as
\[
    IP \simeq \frac{1}{\sqrt{2\pi}} \sum_{j=1}^{n-1}
    e^{-\frac{\sigma_j^2}2}(\sigma_{j+1}-\sigma_j),
\]
where $\{\sigma_j\}_{j=1\,\ldots,\,n}$ are the values of the LOV
parameter corresponding to the impacting LOV orbits.

This case represents another example in which the densification
procedure can be useful: here we do not start from a return with a few
points, but the aim is to obtain a reliable computation of the impact
probability. In particular, to estimate the impact probability of
(29075) in 2880, we convert
$\{\mathbf{y}_1,\mathbf{y}_2,\mathbf{y}_3\}$ in a return with 31
points, 20 of which fall in the Earth impact cross section. The
resulting endpoints are $\sigma_i=1.9962$ and $\sigma_f=1.9988$, and
the impact probability turns out to be $IP = 1.4\times 10^{-4}$, which
is consistent with the value $1.2\times 10^{-4}$ computed by
Sentry. This agreement is particularly remarkable, since the JPL team
performed the impact monitoring with a completely different strategy,
based on a Monte Carlo method \citep{farnocchia:29075}.

\section{Conclusions}

In this paper we presented a densification algorithm to improve the
completeness of the VI search when applying the LOV method. Indeed,
although the LOV sampling proposed in \cite{delvigna:compl_IM} should
guarantee the achievement of a pre-fixed generic completeness level,
in practice VIs with impact probability close to the completeness
level can escape the detection. Typically, this is the case for
returns with very few points, which indicate a strong non-linearity
introduced by previous close approaches.

The idea of our algorithm is to densify returns with length at most
$3$ with a procedure consisting of two steps. The first attempt tries
to obtain returns with $5$ points, where the indices of the new points
are computed according to the structure of the original return. The
addition of a new point requires the selection of its real index, the
interpolation of the LOV at that index and the propagation to the TP
corresponding to the original return. In particular, this last
operation may fail since the LOV can exit the TP around the selected
index, breaking the assumption of simplest geometry. In this case, a
second attempt is performed to add a new point between a successful
and an unsuccessful point of the first attempt. The resulting possibly
densified return is then analysed in the standard way (see
Section~\ref{sec:LOV}). The whole densification process increases the
global computational load of the impact monitoring run: indeed, on one
hand every time the algorithm tries to add a new point a propagation
is performed, and on the other hand longer returns may increase the
application of the iterative methods to search for minimum distance
points. Nevertheless, this is a minor issue thanks to the currently
available computational resources.

The results reported in this paper show that our method has two main
implications in impact monitoring. As the example of 2017~WT$_{28}$
suggests, the densification procedure not only increases the number of
computed VIs, but also allows the detection of VIs with impact
probability above the generic completeness level. This result is
particularly meaningful since it indicates that our densification
method represents a way to fill the gap between the actual
completeness level and the theoretical generic completeness, as
discussed in Section~\ref{sec:motivations}. In particular, a run of
the entire risk list including densification is required to obtain
histograms analogous to those presented in
\cite{delvigna:compl_IM}. With these new histograms it will be
possible to measure the effective improvement of the densification on
the completeness of the impact monitoring problem. This will be
subject of future research.

Another example in which the densification technique turns out to be
useful is represented by asteroid (29075)~1950~DA. This asteroid is
one of the most remarkable cases currently present in the risk lists
of both NEODyS and JPL, since it has a comparatively high probability
of impacting the Earth in 2880 and its hazard assessment involves also
the Yarkovsky effect. As Figure~\ref{fig:tp_nodens} shows, the
original LOV sampling already contains an impacting orbit, so the
detection of the VI is straightforward. Anyway, a local densification
around this orbit allows the addition of a suitable number of points
in the Earth impact cross section, so that we can resort to a
1-dimensional estimate of the impact probability, which is known to be
more accurate when the width is small and the probability is high.

Lastly, it is worth mentioning that the densification algorithm can
applied for a more general purpose. If a return containing impacting
orbits is suitably densified, at least locally in a neighbourhood of
the VI, the determination of the VI representative could be improved
in such a way that the selected orbit is as close as possible to the
VI centre. This turns out to be important when the VI representative
is used as a starting point for further predictions, as it happens for
the semilinear method presented in \cite{dimare:impcor} to compute the
impact corridor of an Earth-impacting asteroid. More general
algorithms for a local and possibly adaptive densification of the LOV
sampling will be subject of future research.

\section*{Acknowledgements}

This work is devoted to the memory of Prof. Andrea Milani
Comparetti. By writing this paper we are keeping a promise: Andrea
gave us the hint for the densification method, which he really cared
about. We would like to thank him a lot for everything he taught us
over the years.

A.~Del Vigna and F.~Guerra acknowledge support by the company
SpaceDyS. This research was conducted under European Space Agency
contract no. 4000123583/18/D/MRP ``P3-NEO-XIII NEODyS Migration
Part~2''.

\bibliographystyle{elsarticle-harv} \bibliography{dens_biblio}

\begin{thebibliography}{21}
\expandafter\ifx\csname natexlab\endcsname\relax\def\natexlab#1{#1}\fi
\expandafter\ifx\csname url\endcsname\relax
  \def\url#1{\texttt{#1}}\fi
\expandafter\ifx\csname urlprefix\endcsname\relax\def\urlprefix{URL }\fi

\bibitem[{{Chesley}(2006)}]{chesley:apophis}
{Chesley}, S.~R., 2006. {Potential impact detection for Near-Earth asteroids:
  the case of 99942 Apophis (2004~MN4)}. In: {Daniela}, L., {Sylvio Ferraz},
  M., {Angel}, F.~J. (Eds.), Asteroids, Comets, Meteors. Vol. 229 of IAU
  Symposium. pp. 215--228.

\bibitem[{{Chesley} et~al.(2014){Chesley}, {Farnocchia}, {Nolan},
  {Vokrouhlick{\'y}}, {Chodas}, {Milani}, {Spoto}, {Rozitis}, {Benner},
  {Bottke}, {Busch}, {Emery}, {Howell}, {Lauretta}, {Margot}, and
  {Taylor}}]{chesley:Bennu}
{Chesley}, S.~R., {Farnocchia}, D., {Nolan}, M.~C., {Vokrouhlick{\'y}}, D.,
  {Chodas}, P.~W., {Milani}, A., {Spoto}, F., {Rozitis}, B., {Benner},
  L.~A.~M., {Bottke}, W.~F., {Busch}, M.~W., {Emery}, J.~P., {Howell}, E.~S.,
  {Lauretta}, D.~S., {Margot}, J.-L., {Taylor}, P.~A., Jun. 2014. {Orbit and
  bulk density of the OSIRIS-REx target Asteroid (101955) Bennu}. Icarus 235,
  5--22.

\bibitem[{Conte and De~Boor(1980)}]{conte-deboor}
Conte, S.~D., De~Boor, C.~W., 1980. Elementary Numerical Analysis: An
  Algorithmic Approach, 3rd Edition. McGraw-Hill Higher Education.

\bibitem[{Del~Vigna et~al.(2018)Del~Vigna, Faggioli, Spoto, Milani, Farnocchia,
  and Carry}]{delvigna:yarko}
Del~Vigna, A., Faggioli, L., Spoto, F., Milani, A., Farnocchia, D., Carry, B.,
  Sep. 2018. {Detecting the Yarkovsky effect among near-Earth asteroids from
  astrometric data}. Astronomy \& Astrophysics 617, A61.

\bibitem[{Del~Vigna et~al.(2019{\natexlab{a}})Del~Vigna, Milani, Spoto, Chessa,
  and Valsecchi}]{delvigna:compl_IM}
Del~Vigna, A., Milani, A., Spoto, F., Chessa, A., Valsecchi, G.~B., Mar.
  2019{\natexlab{a}}. {Completeness of Impact Monitoring}. Icarus 321,
  647--660.

\bibitem[{Del~Vigna et~al.(2019{\natexlab{b}})Del~Vigna, Roa, Farnocchia,
  Micheli, Tholen, Guerra, Spoto, and Valsecchi}]{delvigna:410777}
Del~Vigna, A., Roa, J., Farnocchia, D., Micheli, M., Tholen, D., Guerra, F.,
  Spoto, F., Valsecchi, G.~B., Jul. 2019{\natexlab{b}}. {Yarkovsky effect
  detection and updated impact hazard assessment for near-Earth asteroid
  (410777)~2009~FD}. Astronomy \& Astrophysics 627, A1.

\bibitem[{Dimare et~al.(2020)Dimare, Del~Vigna, Bracali~Cioci, and
  Bernardi}]{dimare:impcor}
Dimare, L., Del~Vigna, A., Bracali~Cioci, D., Bernardi, F., 2020. Use of the
  semilinear method to predict the impact corridor on ground. Celestial
  Mechanics and Dynamical Astronomy 132~(3), 20.

\bibitem[{{Farnocchia} and {Chesley}(2014)}]{farnocchia:29075}
{Farnocchia}, D., {Chesley}, S.~R., Feb. 2014. {Assessment of the 2880 impact
  threat from Asteroid (29075)~1950~DA}. Icarus 229, 321--327.

\bibitem[{Farnocchia and Chesley(2014)}]{farnocchia:1950DA}
Farnocchia, D., Chesley, S.~R., 2014. Assessment of the 2880 impact threat from
  asteroid (29075) 1950 da. Icarus 229, 321--327.

\bibitem[{{Farnocchia} et~al.(2013){Farnocchia}, {Chesley}, {Chodas},
  {Micheli}, {Tholen}, {Milani}, {Elliott}, and
  {Bernardi}}]{farnocchia:apophis}
{Farnocchia}, D., {Chesley}, S.~R., {Chodas}, P.~W., {Micheli}, M., {Tholen},
  D.~J., {Milani}, A., {Elliott}, G.~T., {Bernardi}, F., May 2013.
  {Yarkovsky-driven impact risk analysis for asteroid (99942) Apophis}. Icarus
  224, 192--200.

\bibitem[{{Giorgini} et~al.(2008){Giorgini}, {Benner}, {Ostro}, {Nolan}, and
  {Busch}}]{giorgini:apophis}
{Giorgini}, J.~D., {Benner}, L.~A.~M., {Ostro}, S.~J., {Nolan}, M.~C., {Busch},
  M.~W., Jan. 2008. {Predicting the Earth encounters of (99942) Apophis}.
  Icarus 193, 1--19.

\bibitem[{{Giorgini} et~al.(2002){Giorgini}, {Ostro}, {Benner}, {Chodas},
  {Chesley}, {Hudson}, {Nolan}, {Klemola}, {Standish}, {Jurgens}, {Rose},
  {Chamberlin}, {Yeomans}, and {Margot}}]{giorgini:29075}
{Giorgini}, J.~D., {Ostro}, S.~J., {Benner}, L.~A.~M., {Chodas}, P.~W.,
  {Chesley}, S.~R., {Hudson}, R.~S., {Nolan}, M.~C., {Klemola}, A.~R.,
  {Standish}, E.~M., {Jurgens}, R.~F., {Rose}, R., {Chamberlin}, A.~B.,
  {Yeomans}, D.~K., {Margot}, J.-L., Sep. 2002. {Asteroid 1950 DA's Encounteqr
  with Earth in 2880: Physical Limits of Collision Probability Prediction}. In:
  AAS/Division of Dynamical Astronomy Meeting \#33. Vol.~34 of Bulletin of the
  American Astronomical Society. p. 934.

\bibitem[{Milani et~al.(2005{\natexlab{a}})Milani, Chesley, Sansaturio, Tommei,
  and Valsecchi}]{milani:clomon2}
Milani, A., Chesley, S., Sansaturio, M.~E., Tommei, G., Valsecchi, G.~B.,
  2005{\natexlab{a}}. Nonlinear impact monitoring: line of variation searches
  for impactors. Icarus 173, 362--384.

\bibitem[{{Milani} et~al.(2009){Milani}, {Chesley}, {Sansaturio}, {Bernardi},
  {Valsecchi}, and {Arratia}}]{milani:Bennu_impact}
{Milani}, A., {Chesley}, S.~R., {Sansaturio}, M.~E., {Bernardi}, F.,
  {Valsecchi}, G.~B., {Arratia}, O., Oct. 2009. {Long term impact risk for
  (101955) 1999~RQ$_{36}$}. Icarus 203, 460--471.

\bibitem[{Milani et~al.(1999)Milani, Chesley, and Valsecchi}]{milani:AN10}
Milani, A., Chesley, S.~R., Valsecchi, G.~B., 1999. Close approaches of
  asteroid 1999 an10: resonant and non--resonant returns. Astronomy \&
  Astrophysics 346, L65--L68.

\bibitem[{{Milani} and {Gronchi}(2010)}]{milani:orbdet}
{Milani}, A., {Gronchi}, G.~F., 2010. {Theory of Orbit Determination}.
  Cambridge University Press.

\bibitem[{Milani et~al.(2005{\natexlab{b}})Milani, Sansaturio, Tommei, Arratia,
  and Chesley}]{milani:multsol}
Milani, A., Sansaturio, M., Tommei, G., Arratia, O., Chesley, S.~R., Feb.
  2005{\natexlab{b}}. Multiple solutions for asteroid orbits: {C}omputational
  procedure and applications. Astronomy \& Astrophysics 431, 729--746.

\bibitem[{{Spoto} et~al.(2014){Spoto}, {Milani}, {Farnocchia}, {Chesley},
  {Micheli}, {Valsecchi}, {Perna}, and {Hainaut}}]{spoto:410777}
{Spoto}, F., {Milani}, A., {Farnocchia}, D., {Chesley}, S.~R., {Micheli}, M.,
  {Valsecchi}, G.~B., {Perna}, D., {Hainaut}, O., Dec. 2014. {Nongravitational
  perturbations and virtual impactors: the case of asteroid (410777) 2009 FD}.
  Astronomy \& Astrphysics 572.

\bibitem[{Valsecchi et~al.(2003)Valsecchi, Milani, Gronchi, and
  Chesley}]{valsecchi:resret}
Valsecchi, G.~B., Milani, A., Gronchi, G.~F., Chesley, S.~R., 2003. Resonant
  returns to close approaches: Analytical theory. Astronomy \& Astrophysics
  408, 1179--1196.

\bibitem[{{Vokrouhlick{\'y}} et~al.(2015){Vokrouhlick{\'y}}, {Farnocchia}, {{\v
  C}apek}, {Chesley}, {Pravec}, {Scheirich}, and {M{\"u}ller}}]{vokro:apophis}
{Vokrouhlick{\'y}}, D., {Farnocchia}, D., {{\v C}apek}, D., {Chesley}, S.~R.,
  {Pravec}, P., {Scheirich}, P., {M{\"u}ller}, T.~G., May 2015. {The Yarkovsky
  effect for (99942)~Apophis}. Icarus 252, 277--283.

\bibitem[{Vokrouhlick\'y et~al.(2000)Vokrouhlick\'y, Milani, and
  Chesley}]{vokr:yark}
Vokrouhlick\'y, D., Milani, A., Chesley, S.~R., 2000. Yarkovsky effect on small
  near--earth asteroids: mathematical formulation and examples. Icarus 148,
  118--138.

\end{thebibliography}

\end{document}